\documentclass[conference]{IEEEtran}

\usepackage{picture}
\usepackage{comment}
\usepackage{graphicx,xspace}
\usepackage{epstopdf}
\usepackage{float}
\usepackage{epsfig}
\usepackage{url}
\usepackage{xcolor}
\usepackage{color}

\definecolor{heraldBlue}{rgb}{0.0,0.0,0.8}
\definecolor{heraldRed}{rgb}{0.8,0.0,0.0}
\definecolor{heraldGray}{rgb}{0.4,0.4,0.4}
\definecolor{heraldBlack}{rgb}{0.0,0.0,0.0} 
\definecolor{heraldGreen}{rgb}{0.0,0.4,0.0} 

\newcommand{\ie}{{\em i.e.},\xspace}

\newcommand{\etc}{{\em etc.}\xspace}
\title{Testing for common sense (violation) \\ 
in airline pricing \\ {\Large \emph{or how complexity asymmetry defeated you and the web}}}

\date{}
\begin{document}
\author{\IEEEauthorblockN{Symeon Meichanetzoglou}
    \IEEEauthorblockA{FORTH, Greece\\
    simosme@ics.forth.gr}
    \and
    \IEEEauthorblockN{Sotiris Ioannidis}
    \IEEEauthorblockA{FORTH, Greece\\
    sotiris@ics.forth.gr}
    \and
    \IEEEauthorblockN{Nikolaos Laoutaris}
    \IEEEauthorblockA{Telefonica Research\\
    nikolaos.laoutaris@telefonica.com}
}

\maketitle

\begin{abstract}
We have collected and analysed prices for more than 1.4 million flight tickets involving 
63 destinations and 125 airlines and have found that \emph{common sense violation} 
\ie discrepancies between what consumers would expect and what truly holds 
for those prices, are far more frequent than one would think. 
For example, oftentimes the price of a single leg flight is higher than two-leg 
flights that include it under similar terms of travel 
(class, luggage allowance, \etc). This happened for up to 24.5\% of available 
fares on a specific route in our dataset invalidating the common expectation 
that ``further is more expensive''. 
Likewise, we found several two-leg fares where buying each leg independently 
leads to lower overall cost than buying them together as a single ticket. 
This happened for up to 37\% of available fares on a specific route invalidating 
the common expectation that ``bundling saves money''. Last, several single stop 
tickets in which the two legs were separated by 1-5 days (called multicity fares), 
were oftentimes found to be costing more than corresponding back-to-back fares 
with a small transit time. This was found to be occurring in up to 7.5\% fares 
on a specific route invalidating that ``a short transit is better than a longer one''. 

\end{abstract}

\section{Introduction}
\label{sec:intro}

The so-called {\em information asymmetry} has historically hampered consumers in 
their negotiations with market-savvy sellers. One of the promises of the world-wide-web 
has been to eradicate this information asymmetry for the benefit of consumers. 
For example, price comparison sites allow buyers to easily compare the prices 
of different stores for the same item or service. Similarly, 
consumer and employment fora provide information about once hidden bids for 
real estate~\cite{realtor} or salary levels~\cite{glassdoor},
thus making it easier for individuals to negotiate with sellers.

Traveling, and in particular airline flights, has been one of the sectors 
benefiting the most from the information opennesses offered by the web. 
Where once a traveler was totally dependent upon a travel agent for getting the 
right deal for him, now a customer can have direct access to myriads of travel 
options and real time prices and can thus select on his own. Travelers 
have been delighted by the ability to get access to 
raw flight pricing information coupled with advanced features for searching
it and comparing across different carriers. 

Simultaneously, however, travelers are being exposed to the sheer complexity of 
flight ticket pricing that includes a huge number of factors, 
that when all put together it makes it quite impossible to know when one is 
getting a better deal and when not, despite the existence of readily searchable 
data and advanced search functions.

Indeed, putting together yield management (thus variable pricing),
fares rules, taxation, level of competition, myriads of different flight classes 
and levels within a class, different sellers and meta resellers of tickets, etc.,
creates a number of alternatives that seems to defeat even the most elaborate search facilities. 
From a computational complexity point of view, finding the cheapest fare or 
fares for a specific route is a hard problem for the online search engines and the travelers\cite{compFares}.
 
Therefore, it seems that although the information asymmetry barrier has 
been lifted, a new \emph{complexity asymmetry} barrier has promptly replaced it.
Travelers are overloaded with the sheer amount of options offered to them and 
thus, in the end, are rather doubtful about whether they are getting the best possible deal or not.
 
Their confusion is sometimes caused by observed violations of what most 
consider, or take for granted, as a ``common sense'' rule in flight ticket pricing. 
Take for example the following three common sense rules:
 
\begin{enumerate}
\item Common Sense 1 (CS1). ``A multi-leg ticket should cost more than its 
    individual legs''. The cost of the multi-leg ticket is obviously higher than any of its legs. 
A violation of this common sense is usually called ``hidden city ticketing''.

\item Common Sense 2 (CS2). ``The price of a single stop flight should be lower 
    than bying the individual legs separately''. ``Bundling'' is supposed to save money for the consumer.

\item Common Sense 3 (CS3). ``If customer convenience is driving price then 
    short transit (e.g. 1-2 hours) should cost more than split by day flights 
    (second leg after 1 or more days)''. 
Customers would rather do a quick transit rather than spend a day at an airport 
hotel or even worse at waiting area or lounge.

\end{enumerate}

There are anecdotal rumors on violations of the above common sense rules
\cite{howToBeatHighAirfares},\cite{techDirtArticle}. 
But still, it is widely unknown how frequently such violations occur, in which 
routes, which airlines, and when. Interestingly, there is more awareness 
as to why such violations occur (competition in pricing over different routes, 
computational complexity of pricing algorithms) 
as opposed to the quantitative questions posed above.

\noindent \textbf{Our contribution:} In this paper we present a quantitative study
on the above mentioned questions about CS violations in a dataset containing
tickets in routes in Europe. Using an online search engine for airline fares,
we retrieved around 1.4 million tickets and performed an analysis to identify
and calculate the violations percentages for each violation presented above.

\noindent \textbf{Summary of results:}
\begin{itemize}

\item Overall (across all routes), 1.53\% of the 
    single stop fares have CS1 violations. It is clear that the violations of 
    CS1 do not happen frequently across the dataset. 
However, for some particular routes, there is a relatively high percentage of 
violations. The BRU-STR route has a 24.5\% violations. 
Similarly, specific airlines have a large number of violations. 
A Dutch airline has 565 violations
in 1055 single stop tickets in our dataset. 
A Scandinavian airline has 209 violations in 4659 single stop
tickets.

\item Overall, 1.99\% of the single stop fares 
    have CS2 violations. 
Again, despite the low overall frequency across the 63 routes of the dataset, 
in specific routes the violations have high frequency.
In ORY-TLS CS2 violations reached 39.3\%. 

\item Overall 5.75\% of the single stop tickets 
    in the dataset have CS3 violations. 
The biggest percentage is in OSL-CPH route, with 26.7\% violations. 
\end{itemize}

Please note that our objective \underline{is not} to explain \emph{why} 
these violations occur, but \emph{when} and \emph{where} 
they occur and by \emph{which} airlines. 
The \emph{why}, if at all possible to answer, 
would require deep insight into the way that airlines operate, which 
is clearly beyond our capacity and intention.

\section{Dataset}
\label{sec:experiment}

To gather our dataset of airline tickets for analysis, 
we queried \url{matrix.itasoftware.com}, an online search engine
for airfares. Matrix uses a database of flights, prices, and seat
availability that is updated by the private networks of most of the airlines of the world. 
Based on a user provided query for a trip, it returns all the available options, \ie
all available fares that match the route and travel dates. 
It also returns detailed information about the fares, booking codes and the rules that
apply.

In order to have a balanced dataset, we chose routes that span a variety of Hub and Regional airports.
We classify an airport as Hub if it is in the 20 most busy airports of 2013
according to Wikipedia\footnote{\url{https://en.wikipedia.org/wiki/List_of_the_busiest_airports_in_Europe}}. 
Otherwise we classify it as non-Hub (Regional).

\subsection{Dataset description}
In total the data contain 1404942 tickets across 63 routes in Europe for departing dates 
across 2014 and 2015. The crawling of the itasoftware service
and the collection of the tickets tool place in 2014
Out of the 63 routes, 20 were between hub airports, 20 between hub and regional airports
and the remaining 23 between regional. The selection of the airports was done
at random from the list of the busiest airports in Europe. For example,
for a hub-to-hub route, we chose two random airports from the 20 most busy airports
(as explained above). 

We collected three different types of fares
\begin{itemize}
\item Single stop. Tickets that comprise two flights (legs).
\item Separate legs. Tickets that are direct and correspond to one leg of 
a single stop ticket. Specifically, after having retrieved a single stop
ticket, we query for two direct tickets that correspond to each one of 
the legs of the single stop ticket.
\item Multicity. Similar to single stop but the date of the second leg
is specified by the user. In our case, we query for tickets where the
date of the second leg is after 1-5 days from the first.
\end{itemize}

The distribution of the tickets in the different types is shown in Table~\ref{tab:ticketsDist}.
 \begin{table}[H]
\centering
\caption{The number of collected tickets per category}
\begin{tabular}{l*{10}{c}}
Single stop & Separate legs & Multicity \\
\hline
129105 & 449919 & 825918
\end{tabular}
\label{tab:ticketsDist}
\end{table}

\subsection{Data collection}
Although \url{matrix.itasoftware.com} does not offer an API,
all the data passed between the browser and the service are JSON formatted. Thus,
we were able to write a python program that queries the service, parses the JSON response 
and stores the data in a database. 

\begin{figure*}[t]
\centering
\includegraphics[width=\textwidth]{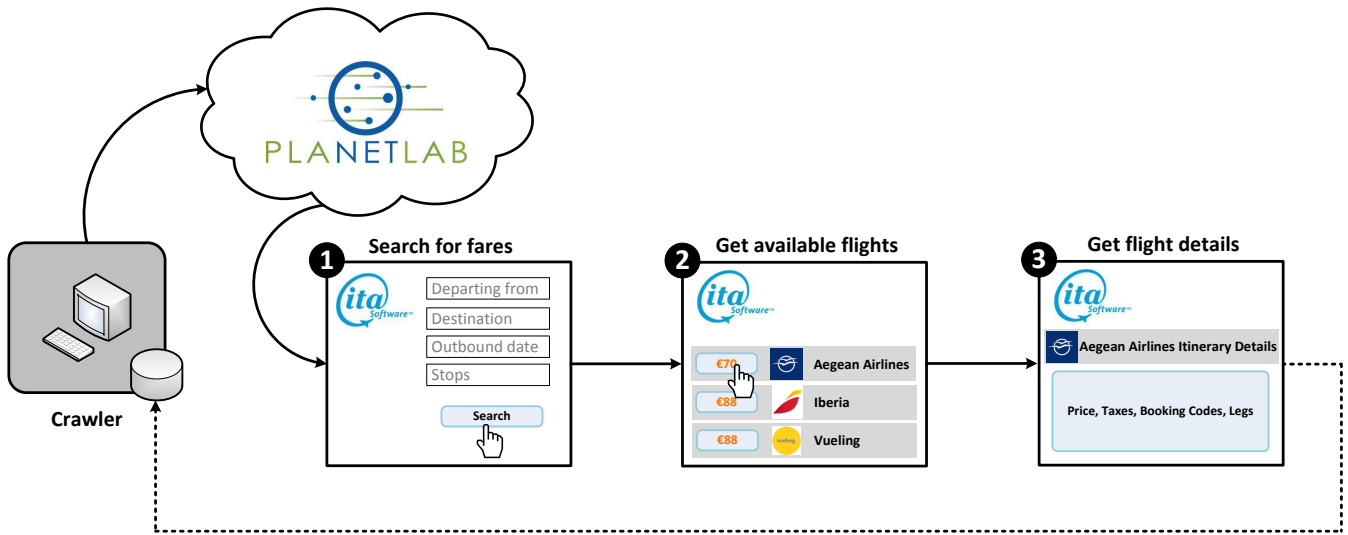}
\caption{The architecture of the crawler}
\label{fig:crawler}
\end{figure*}

Our crawler queries \url{matrix.itasoftware.com} and collects data about airfares,
using the PlanetLab infrastructure and the SOCKS protocol. 
Each PlanetLab node acts as a SOCKS proxy and each request is tunneled through a proxy chosen randomly. 
The steps are shown in Fig.~\ref{fig:crawler}
\begin{itemize}
    \item In step 1, we make a request for fares for a specific route, date and 
    type of ticket (direct, single stop, multicity).
    \item In step 2, we get the list of available fares.
    \item In step 3, we request the details for each fare returned in step 2. 
    These details include: fare price, list of taxes and their prices, list of legs, list of booking 
    codes. All data are stored in a MySQL database.
\end{itemize}

The crawling starts by quering for single stop tickets (two legs). 
The middle stop airport is saved and, subsequently, the crawler queries for direct tickets: 
depart airport - middle stop and middle stop - arrival airport. 
Thus, for each single stop ticket, we query the price of each one of its legs as if it were being booked 
as a single direct flight. Finally, we query for \emph{multicity} fares.

We used MySQL to store the collected information.

\section{Analysis}
\label{sec:analysis}
Next, we present our quantitative analysis of the three types of common sense violation that we have defined on the introduction.
We make sure that any fares that are compared have (1) the same booking code 
(T, Y, X, etc.), (2) are offered by the same airline, and (3) are collected
during the same day. 

Each ticket price consists of two parts: the fare price and the taxes.
Thus, when two tickets are found to have different prices any of these components
might have been the reason. For example, the taxes might be identical but the fare prices
might be different. In the following sections, we present all violations
(caused by either different fare prices or taxes) as well as those that are guaranteed to be only due to fare prices.


\subsection{Common Sense 1}
\label{subsec:cs1}
Common Sense 1 compares the price of entire single stop tickets with the prices 
of its separate legs when booked as independent direct flights. 
For each single stop ticket, we retrieve the equivalent direct fares for each leg 
of the single stop ticket. 
To be considered equivalent, such a direct fare must have the same flight number 
and booking code (T, Y, X, etc.) with the respective leg flight number of the 
single stop ticket.\footnote{However, in some cases, airlines do not offer the same booking codes for 
both the single stop fare and the separate leg fare. 
A possible explanation of this is the seat availability management done by the airlines.}

If any of the respective direct tickets  is more expensive than the relevant 
single stop ticket then we have a violation of CS1 and count it. 

\begin{figure}[H]
\centering
\includegraphics[width=\columnwidth]{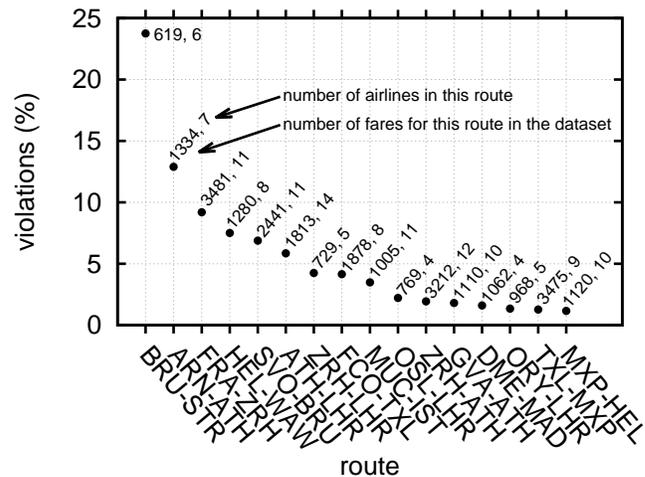}
\caption{Percentages of the tickets that violate common sense 1. 
The labels indicate the number of fares per route and the number of airlines that 
have fares in each route}
\label{fig:cs1Violations}
\end{figure}

In Fig.~\ref{fig:cs1Violations} we show the violations percentages for each route. 
The labels indicate the number of fares for each route, and the violations percentage
(y-axis) indicate the portion of these fares that violate CS1. Also, 
in the labels we show the number of airlines that have fares in each route.
All CS1 violations in our dataset are caused by different fare prices. This
is in contrast to CS2,3 where there are violations caused by different taxes 
(section 3.2).

As shown in Fig.~\ref{fig:cs1Violations} there are 6 routes out of the 64 that 
have a CS1 violations percentage above 5\%. Overall half of the routes (32) 
had at least one violation. 
Although the CS1 violations are not so common across the entire dataset (1.53\%), when 
they are observed on a particular route, their frequency can be high, as in 
BRU-STR \footnote{\url{www.world-airport-codes.com}} where it reaches 24\%. 
To investigate what happens in the BRU-STR route 
we plot the breakdown of CS1 violations per route and airline. 
We intend to see is whether a high percentage of CS1 violation on a route is 
attributed to several of the airlines operating on the route, or it is the 
result of the pricing policy of one or few airlines. 

\begin{figure*}[t]
\centering
\includegraphics[width=\textwidth]{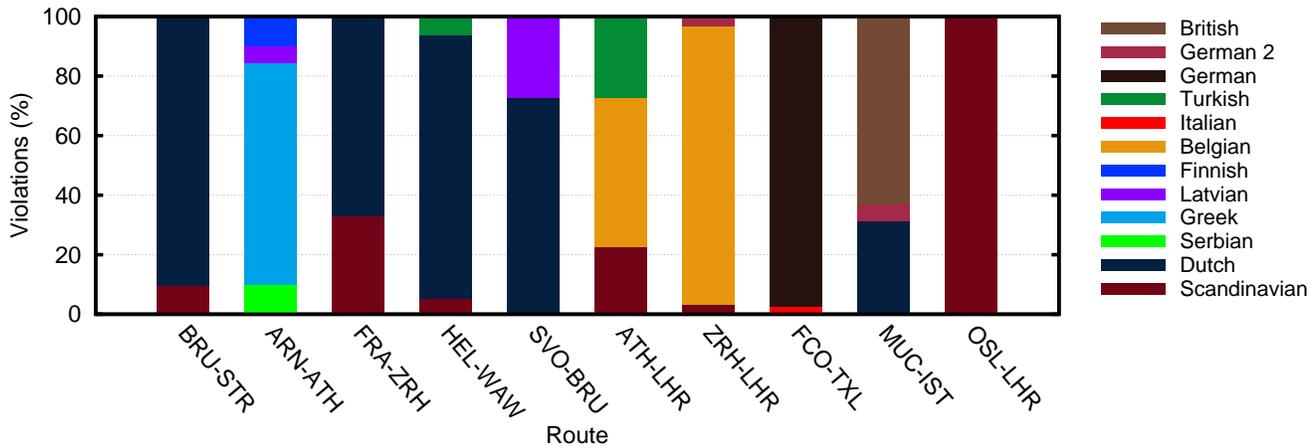}
\caption{Percentages of the violations per airline and route}
\label{fig:cs1PerRoutePerAirline}
\end{figure*}

We do this in Fig.~\ref{fig:cs1PerRoutePerAirline} from which we conclude that, usually, 
one or two airlines cause the majority of the violations on a route with high 
frequency of CS1 violations.  This means that the violations are a result of the
policies of particular airlines rather than a result of the route itself.  
From Fig.~\ref{fig:cs1PerRoutePerAirline} it is clear that for FRA-ZRH, BRU-STR,
HEL-WAW the majority of violations come from just two airlines (the Scandinavian airline, the Dutch airline).

\begin{figure}[H]
\centering
\includegraphics[width=\columnwidth]{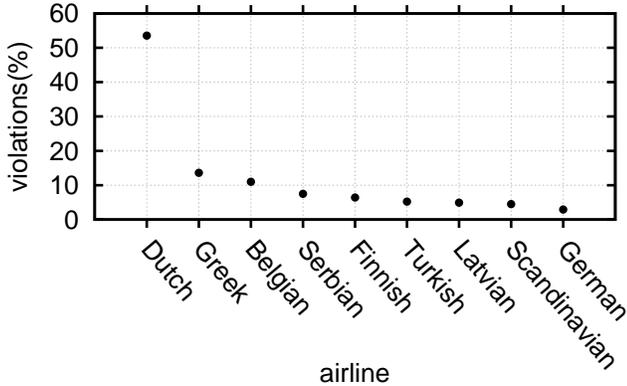}
\caption{Percentages of the tickets per airline that violate common sense 1}
\label{fig:cs1PerAirline}
\end{figure}

We now turn into re-examining the data per airline instead of per route. 
In Fig.~\ref{fig:cs1PerAirline} we show the percentages of violations
for each airline. For each airline these violations might be spread across
one or many routes. We examine this distribution in the next section.

We see that the Dutch airline has an exceptionally high percentage of violations (53.5\%). 
The Greek airline and the Belgian airline are around 10\% and most others are below 10\%.
In absolute numbers, the Dutch airline has the most violations (565) with the Scandinavian airline coming second (209).
In the next section we focus in the airlines with the most violations, and we
research the distribution of the violations in different routes.

\subsubsection{The airlines with the highest number of CS1 violations}
As shown in Fig.~\ref{fig:cs1PerAirline}, the Dutch airline has the highest percentage of 
violations on the fares that it offers (53.5\%). The Greek airline comes second with 13.6\% of its 
single stop fares violating CS1.
The distribution of the violations in the routes that the Dutch airline offers is shown in Table~\ref{tab:KLM}.
The violations happen in five routes. As shown, most violations of the Dutch airline (37.9\%) happen on FRA-ZRH.
Furthermore, for the first four (FRA-ZRH, BRU-STR, SVO-BRU, HEL-WAW), 
the Dutch airline causes the vast majority of the violations for these routes, 
as shown in Fig.~\ref{fig:cs1PerRoutePerAirline}.

\begin{figure}[H]
\centering
\includegraphics[width=\columnwidth]{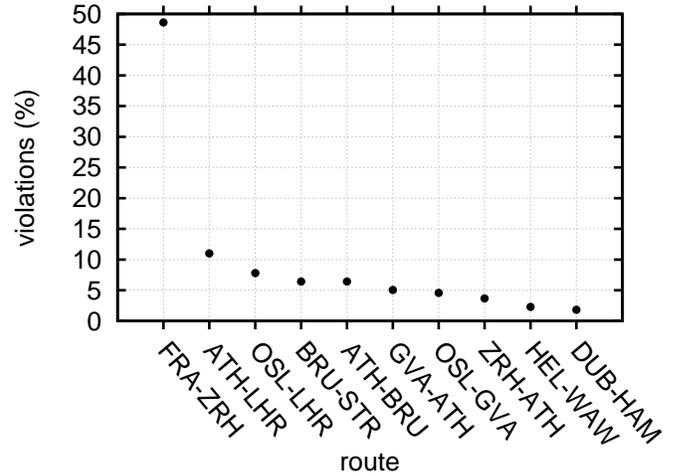}
\caption{Distribution of the violations per route for the Scandinavian airline}
\label{fig:The Scandinavian airline}
\end{figure}

\begin{table}[H]
\caption{Distribution of the violations per route for the Dutch airline}
\centering
\begin{tabular}{l*{10}{c}}
Route & Violations (\%) \\
\hline
FRA-ZRH & 37.88 \\
BRU-STR & 23.54 \\
SVO-BRU & 21.6 \\
HEL-WAW & 15.04 \\
MUC-IST & 1.95 \\
\end{tabular}
\label{tab:KLM}
\end{table}
The Scandinavian airline is second in absolute number of violations (218 violations). 
The distribution of violations is shown in Fig.~\ref{fig:The Scandinavian airline}. 
Almost half of the violations (48.62\%) that the Scandinavian airline has in its fares take place in FRA-ZRH. 
In total, the Scandinavian airline has CS1 violations in ten routes.

Based on the above graphs, it is clear that both airlines seem to have a 
special policy for FRA-ZRH. The majority of the violations in this route are by the Dutch airline and the Scandinavian airline.

As shown in Fig.~\ref{fig:cs1PerAirline}, the Greek airline has the second largest
percentage of CS1 violations (13.6\%). 
In the dataset, the Greek airline has all its CS1 violations in the ARN-ATH route.

A similar situation exists for the Belgian airline: half of the violations take place in 
ATH-LHR. Three more routes follow (ZRH-LHR, GVA-ATH, ZRH-MUC) with much lower frequencies.

We conclude that the violating airlines have their violations 
distributed in a number of routes (from one to ten in the case of the Scandinavian airline).
However, in each airline there is a single route where the majority (40-50\%)
of the airline's violations take place.


\subsection{Common Sense 2}
\label{subsec:cs2}

\begin{figure}[H]
\centering
\includegraphics[width=\columnwidth]{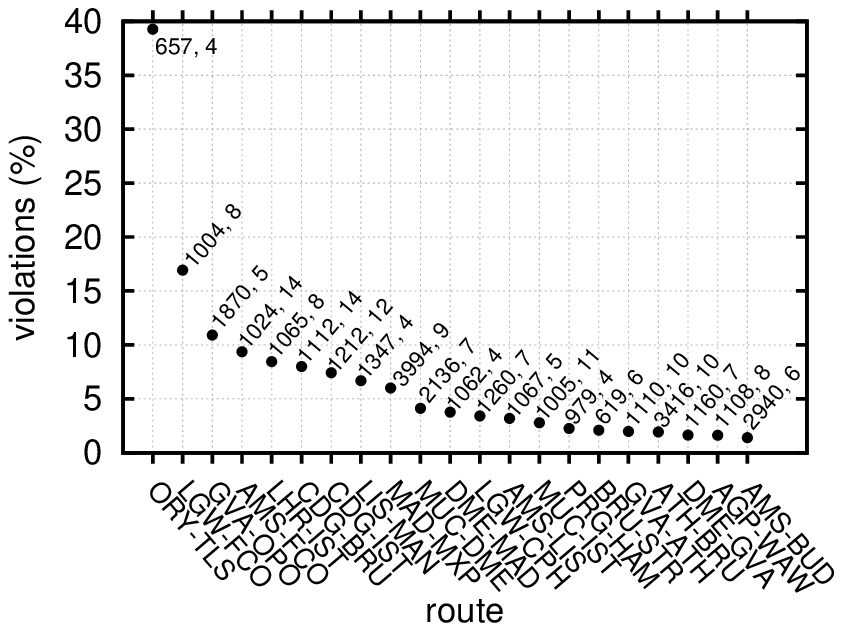}
\caption{Percentages of the tickets that violate common sense 2.
The lables indicate the number of fares for each route in the dataset, and
the number of airlines that have fares in each route.}
\label{fig:cs2Violations}
\end{figure}

Here, we derive all combinations of the available tickets for the first leg with the available tickets 
for the second leg, to create a two legs ticket equivalent to the original single stop ticket. 
If this ``created" ticket is more expensive than the single stop ticket then we 
have a violation of CS2. The violations could be caused by differences in fare prices,
or different taxes between the compared tickets \footnote{The different taxes are imposed
by the airports.} In contrast to CS1, the CS2 violations are caused by both
different prices and different taxes.
To illustrate this difference, we show the violations percentages per route in 
both the case of violations caused by different full prices (\ie different taxes, different
fare prices, or both) and by different fare prices alone. 
The full price violations of CS2 are shown in Fig.~\ref{fig:cs2Violations}.
There are nine routes that have violations above 5\%. In one case (ORY-TLS)
the violations percentage reaches 39.3\%. In total, there are 21 routes
that have violations above 1\%. However, most of them (18) have percentages
below 10\%. As in CS1, although that the overall violation frequency is low 
(1.99\%), for specific routes the violations percentage is significant.

\begin{figure}[H]
\centering
\includegraphics[width=\columnwidth]{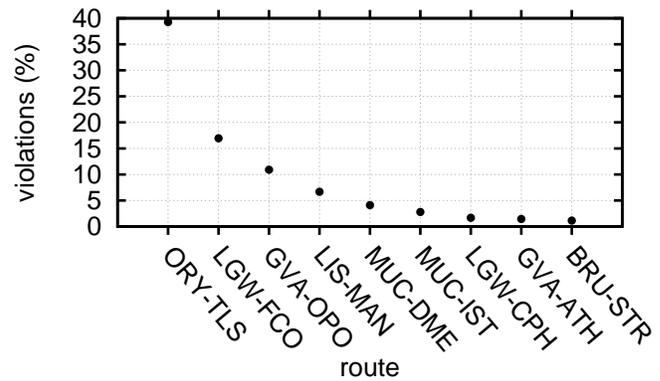}
\caption{Percentages of the tickets per route that violate common sense 2 caused by 
different fare prices}
\label{fig:cs2FaresViolations}
\end{figure}

The violations caused by different fare prices only are shown in 
Fig.~\ref{fig:cs2FaresViolations}
In this plot we have excluded the violations that are caused by different taxes
\ie violations that originate from airports policies\footnote{As described in the dataset description,
our dataset contains detailed information about the fare price and the prices of the taxes for each ticket.}. 
As can be seen, only nine 
airlines have violations here. Still, the top three airlines that had violations
percentags above 10\% in Fig.~\ref{fig:cs2Violations} have the same number of violations
as in Fig.~\ref{fig:cs2FaresViolations}.

\begin{figure}[H]
\centering
\includegraphics[width=\columnwidth]{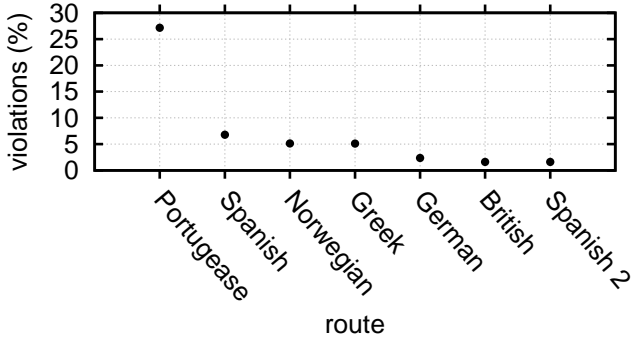}
\caption{Percentages of the fare prices of specific airlines that violate
common sense 2. In this plot are shown only the violations that are caused
by differences in the fare prices (we excluded violations caused by different
taxes).}
\label{fig:cs2FaresPerAirline}
\end{figure}

In the remaining part of this section we will focus on the violations that are caused
by different fare prices only. The reason is that we strive to capture possible 
airlines policies that lead to CS2 violations, and not violations that are caused
by different taxes imposed by airports or states. Along these lines, 
in Fig.~\ref{fig:cs2FaresPerAirline} we depict the airlines with CS2 violations
percentage above 1\%.

\begin{figure*}[t]
\centering
\includegraphics[width=\textwidth]{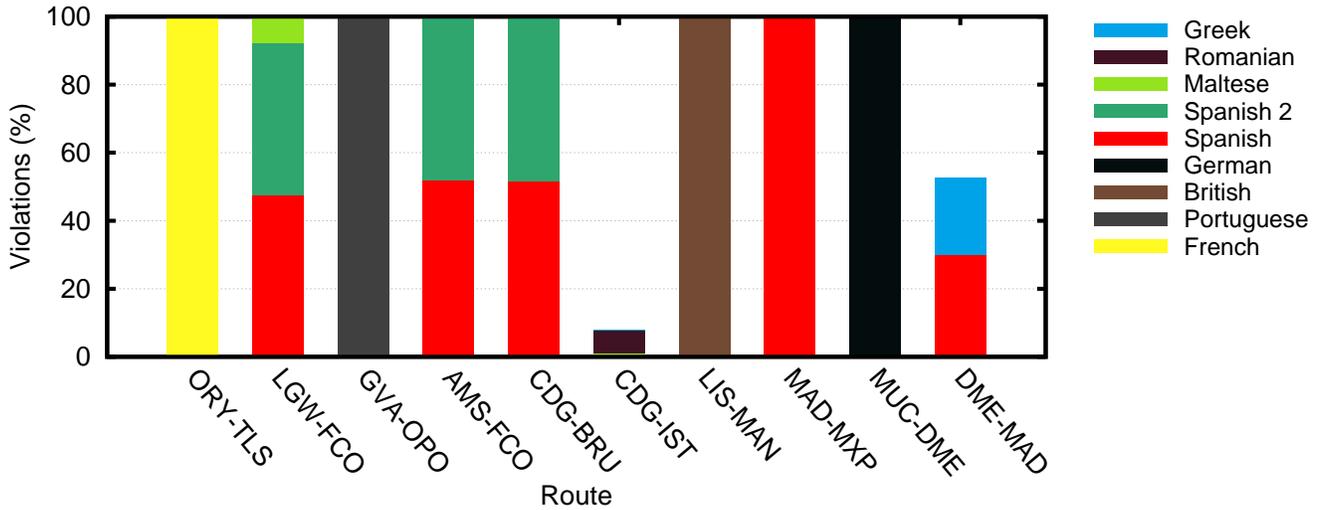}
\caption{Percentages of violations per airline and route. The bars that
do not sum up to 100\% indicate that there are violations that are 
caused by single stop fares that are offered by different airlines.}
\label{fig:cs2PerRouteAirline}
\end{figure*}

To find out whether the airlines violate CS2 in one or many routes, in Fig.~\ref{fig:cs2PerRouteAirline}
we depict the distribution of violating airlines per route.
Similarly to CS1 violations, usually a single or two airlines are responsible for the majority of
CS2 violations in a route.
This means that the violations must relate to the pricing policies of these airlines 
and not so much on the routes themselves. 

Next, we focus exclusively on the airlines and their violations.
In Fig.~\ref{fig:cs2FaresPerAirline} we depict the airlines with the 
highest frequency of CS2 violations. 

We see that there are four airlines with violations above 5\% out of seven 
airlines that have violations in our dataset. In one case,
(Portuguese airline) the violations reach 27.16\%.

\subsubsection{The airlines with the highest number of CS2 violations}
The Portuguese airline has the highest percentage of CS2 violations, at 27.16\%.
All violations take place in GVA-OPO route, as shown in Fig.~\ref{fig:cs2PerRouteAirline}.
The Spanish airline has 6.8\% violations spread across a number of routes. 
The majority of the violations take place in LGW-FCO (81 violations)
and a few violations take place in CDG-BRU (3 violations).
Similarly, the Norwegian airline has all violations in LGW-CPH (21 violations).

The above distributions indicate that in CS2, the top violating airlines have the vast majority
of their violations in one route. This is in contrast to CS1, where the top CS1 violating
airlines have violations across a number of routes. Although in CS1 we see that
one route has most violations, there is a number of other routes where the
violations are distributed. Here, the CS2 violations are much more concentrated:
in most cases they take place in a single route for each airline.


\subsection{Common sense 3}
\label{subsec:cs3}
Here we focus on single stop tickets where the second leg comes after 1-5 days
from the date of the first leg (these tickets are called multicity tickets).
We compare the prices of these tickets with the equivalent single stop tickets where
the second leg is on the same day as the first leg.
Normally, we expect the multicity fare to be cheaper, since users do not
want to have the delays in transit that the multicity tickets have.
If the multicity ticket is more expensive we count it as a CS3 violation.

After analyzing the dataset, we found that most violations are caused by different
taxes between the compared tickets: in the multicity tickets, some airports impose
extra taxes compared to the single stop fares where the second leg is on the same 
day as the first leg.
For example, in the UK, there is the \emph{United Kingdom Air Passenger Duty} tax, 
which is included only in the tickets where the second leg is after 24 hours from the first.

When focusing on the airlines, most violating airlines have violations in a small
number of routes. For most routes, all violations are caused by one airline. This
indicates that each airline focuses on specific routes and causes all violations
in these routes.

\begin{figure}[H]
\centering
\includegraphics[width=\columnwidth]{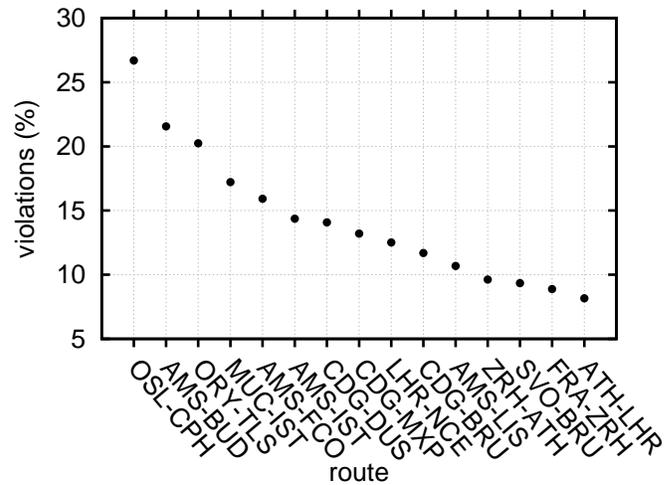}
\caption{Percentages of the tickets that violate common sense 3. These are violations
on the full prices of the tickets. We show only the top-15 most violating routes}
\label{fig:cs5Violations}
\end{figure}

In Fig.~\ref{fig:cs5Violations} we depict the CS3 violations on the full prices
of the tickets. There are 11 routes which have violations above 10\%. Thus, CS3 seems to be 
much more common compared to CS1 \& 2. 

\begin{figure}[H]
\centering
\includegraphics[width=\columnwidth]{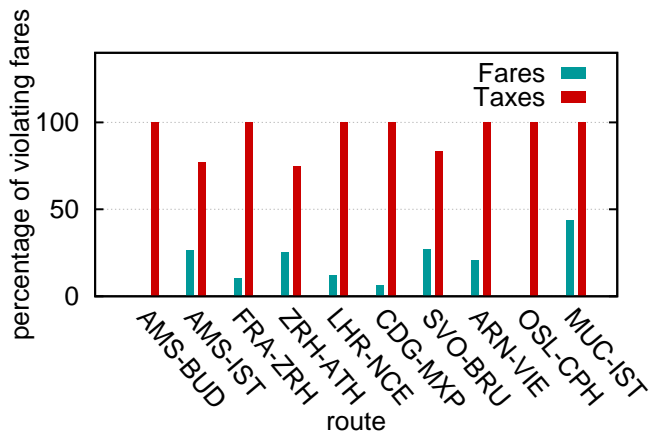}
\caption{Percentages of the violating fares of CS3 that have different fare prices and taxes.
The tickets can have different fare prices, different taxes or both. Thus, the percentages
do not sum up to 100\%}
\label{fig:cs3TaxesOrFares}
\end{figure}

As described previously, most CS3 violations are caused by different taxes imposed 
by the airports. This is shown in Fig.~\ref{fig:cs3TaxesOrFares} where, for the ten
most violated (in absolute numbers) routes, we plot the percentages of the violations
that have different taxes and different fare prices. 

\begin{figure}[H]
\centering
\includegraphics[width=\columnwidth]{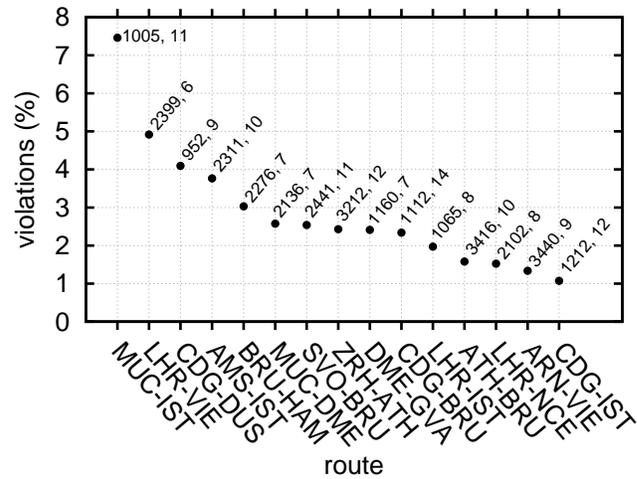}
\caption{Percentages of violations caused by different fare prices.
The labels indicaate the number of fares for each route in the dataset,
and the number of airlines that have fares in each route.}
\label{fig:cs3Fares}
\end{figure}

In Fig.~\ref{fig:cs3Fares} we depict routes that have CS3 violations caused strictly by fare price. 

Compared with CS1,2 violations, it is clear that fewer routes have 
CS3 violations caused only by different fare prices. 
There is only one route that has a percentage above 5\%. Furthermore, 
there are only fifteen routes that have a percentage above 1\%.

\begin{figure}[H]
\centering
\includegraphics[width=\columnwidth]{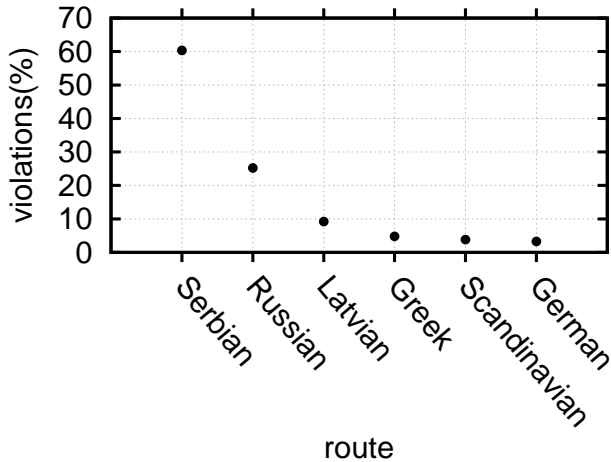}
\caption{Percentages of the tickets per airline that violate common sense 3}
\label{fig:cs3PerAirline}
\end{figure}

Having excluded the violations that are caused by taxes (\ie by the airports 
policies) we can further investigate the violations and identify the airlines
that cause them. In Fig.~\ref{fig:cs3PerAirline}, the CS3 violations per airline are shown.

The Serbian airline has the highest percentage of CS3 violations (60.4\%). Next is the Russian airline
with 25.2\% and the Latvian airline with 9.2\%. In absolute numbers, the most violations are with 
Scandinavian airline (177).

\subsubsection{The airlines with the highest number of CS3 violations}
The Serbian airline has the biggest violations percentage with 60.36\% violations.
(137 violations over 227 single stop tickets). 
In Fig.~\ref{tab:serbia} we show the distribution of violations for the Serbian airline. 

\begin{table}[H]                          
\centering
\caption{Percentages of violations per route for the Serbian airline}
\begin{tabular}{l*{10}{c}}
Route & Violations (\%) \\
\hline
ZRH-ATH & 45.99 \\
ATH-BRU & 45.26 \\
ARN-ATH & 5.11 \\
\end{tabular}
\label{tab:serbia}
\end{table}

The majority of the violations take place in two routes. For ZRH-ATH
and ATH-BRU, the Serbian airline has all violations in these routes. For
ARN-ATH, violations are both from the Serbian airline and the Latvian airline. 

The Russian airline is second with a 25.2\% violations (126 over 500 single stop
tickets). The distribution of the violations in different routes for the Russian
airline is shown in Table~\ref{tab:russian}.
\begin{table}[H]                          
\centering
\caption{Percentages of violations per route for the Russian airline}
\begin{tabular}{l*{10}{c}}
Route & Violations (\%) \\
\hline
AMS-IST & 61.1 \\
MUC-DME & 14.3 \\
LHR-IST & 15.1 \\
CDG-IST & 9.5  \\
\end{tabular}
\label{tab:russian}
\end{table}

For MUC-DME the violations come from both the Russian airline and the Latvian airline.
For all other routes, all violations come from the Russian airline.

The Latvian airline has a 9.2\% violations percentage (68 over 739 fares). The distribution 
of the violations is shown in Table~\ref{tab:baltic}.
\begin{table}[H]                          
\centering
\caption{Percentages of violations per route for the Latvian airline}
\begin{tabular}{l*{10}{c}}
Route & Violations (\%) \\
\hline
ARN-VIE & 76.7 \\
MUC-DME & 16.7 \\
ARN-ATH & 6.7 \\
\end{tabular}
\label{tab:baltic}
\end{table}
Most violations of the Latvian airline take place in ARN-VIE. In this route,
all violations are from the Latvian airline. Air Baltic also has some violations 
in MUC-DME, ARN-ATH, where the Russian airline and the Serbian airline have violations.

\section{Conclusion}
We performed an analysis of airline fare prices
in a dataset of  1.4 million airline tickets collected from matrix.itasoftware.com.
We measured the frequencies of three commmon sense violations (on single stop 
and multicity fares), and found out that, although overall the percentages are small 
(1.53\% for CS1, 1.99\% for CS2, 5.75\% for CS3) when focusing on specific routes
and specific airlines, the violation frequencies can be high.

Furthermore, it seems that the violations originate from the pricing policies of 
the specific airlines, rather than imposed by the route. This is illustrated by
the fact that for each route, one or two airlines cause all violations.

\bibliographystyle{plain}
\bibliography{bibliography}

\end{document}